\documentclass{PoS}

\usepackage{amsmath,amssymb}
\usepackage{verbatim}
\usepackage{graphicx}
\usepackage{color}
\usepackage{footnote}
\usepackage{enumitem}

\newcounter{rowcount}
\usepackage{etoolbox}
\makeatletter
\patchcmd{\chapter}{\if@openright\cleardoublepage\else\clearpage\fi}{}{}{}
\makeatother

\usepackage{textcomp}

\usepackage[stable]{footmisc}
\newcommand{\be}{\begin{eqnarray}}
\newcommand{\ee}{\end{eqnarray}}
\newcommand{\bea}{\begin{eqnarray}}

\newcommand{\eea}{\end{eqnarray}}

\title{Asymptotic symmetry algebras of conformal gravity in four dimensions }

\ShortTitle{Asymptotic symmetry algebras of conformal gravity in four dimensions }

\author{{I. Lovrekovic*}\\
  Technische Universit\"at Wien, Institute for Theoretical Physics\\
        E-mail: \email{lovrekovic@hep.itp.tuwien.ac.at}
}%\footnote{ Thanks to Daniel Grumiller, Robert McNees, Florian Preis, Maria Irakleidou}

\abstract{We consider conformal gravity boundary conditions and outline the highest dimensional non-trivial asymptotic symmetry algebras of conformal gravity. The highest among them is five dimensional and leads to a global geon solution. \\
\\
\\
\\
\\
\\
\\
\\
\\
\\
\\
\\
\\
\\
\\ 
\\
\\
\\
\\
\\
\\
\\
\\
*Thanks to Daniel Grumiller, Robert McNees, Florian Preis, Maria Irakleidou
}

\FullConference{ Corfu Summer Institute 2016 "School and Workshops on Elementary Particle Physics and Gravity"\\
		31 August - 23 September, 2016}

\begin{document}

\section{Introduction and Motivation}
%In the talk we consider the classification of the asymptotic symmetry algebras of conformal gravity. 
In this talk, we consider the asymptotic symmetry algebra (ASA) of conformal gravity (CG).
ASA is generally investigated for three main reasons. First one is to learn about the ASA itself, the second one is to investigate the theory of gravity, and the third one is to learn about the charges defined by the ASA which describe the field theory at the boundary. 
CG is a very interesting theory of gravity that has recently received much attention. Compared to Einstein gravity, it has an advantage that it is two loop renormalizable while Einstein gravity  is not, however, it introduces an issue common for higher derivative gravity theories. It contains ghosts. %Therefore, one can consider it as a toy model.

It was studied from theoretical aspects by Maldacena, who simultaneously showed the importance of the boundary conditions obtaining Einstein gravity solutions from conformal gravity by imposing appropriate boundary conditions \cite{Maldacena:2011mk}. Further motivation to study CG comes from the  the number of papers by 't Hooft in which he considers CG and conformal symmetry, and speculates that conformal symmetry could play a crucial role for the physics at the Planck energy scale \cite{Hooft:2009ms,Hooft:2010ac,Hooft:2010nc,Hooft:2014daa}. CG arises from five dimensional Einstein gravity (EG) as a boundary counterterm %\cite{} 
and from the twistor string theory \cite{Berkovits:2004jj}.  The analysis of the holographic \cite{Grumiller:2013mxa} and canonical \cite{Irakleidou:2014vla} aspect of conformal gravity,  showed that consistent set of boundary conditions leads to well defined variational principle and finite charges of the conformal gravity holography in four dimensions, while the charge associated to Weyl transformations vanishes.
On phenomenological grounds, it was mostly studied by Manheim, in the explanation of the galactic rotation curves without the addition of dark matter \cite{Mannheim:2010xw,Mannheim:2011ds,Mannheim:2012qw}.

The aspect from which we are considering CG is the AdS/CFT correspondence that has been demonstrated to work on number of examples such as $AdS_3/LCFT_2$ \cite{Grumiller:2009mw, Grumiller:2009sn}, AdS/Ricci flat correspondence and other examples of gauge/gravity correspondence. %, and in the framework of the AdS/Ricci flat correspondence. % in which one can apply it on the \textcolor{blue}{black branes black strings, black holes}.

Within this classification, beside ASAs of CG, we also obtain the asymptotic solutions that can be uplifted to the global solutions. These solutions are pp waves or geons \cite{Irakleidou:2016xot}. The classification also contains the known CG solutions, such as Manheim--Kazanas--Riegert solution and rotating black hole solutions \cite{Mannheim:1988dj}.

\section{Conformal gravity}
Given a manifold $\mathcal{M}$, conformal gravity action is described by the 
\begin{equation}
S=\alpha\int d^4x C^{\mu}{}_{\nu\sigma\rho}C_{\mu}{}^{\nu\sigma\rho}\label{accg}
\end{equation}
living on that manifold. 
In (\ref{accg}), $\alpha$ is dimensionless coupling constant responsible for the power counting renormalizability of the action and $C^{\mu}{}_{\nu\sigma\rho}$ is Weyl tensor. %  naturally symmetric under Weyl rescalings of the metri. % while it consists of the traceless part of the Riemann tensor.
 The action is invariant under Weyl rescalings of the metric 
\begin{equation}
g_{\mu\nu}\rightarrow \tilde{g}_{\mu\nu}=e^{2\omega}g_{\mu\nu}
\end{equation}
for $\omega$ Weyl factor. Variation of the action leads to the CG equations of motion  which require vanishing of the Bach tensor
\begin{equation}
\left( \nabla^{\rho}\nabla_{\sigma} +\frac{1}{2}R^{\rho}{}_{\sigma}\right)C^{\sigma}{}_{\mu\rho\nu}=0.\label{bach}
\end{equation}

%\begin{equation}
%S=\alpha\int d^4x C^{\alpha}{}_{\beta\gamma\delta}C_{\alpha}{}^{\beta\gamma\delta}\label{action}
%\end{equation}
%is defined by the square of the Weyl tenors $C_{\alpha\beta\gamma\delta}$ which is traceless part of the Riemann tenosor

\section{Boundary Conditions}

%The boundary conditions that
The asymptotic $(0<\rho<<\ell)$ line element is described with 
\begin{equation}
ds^2=\frac{\ell^2}{\rho^2}\left(-\sigma d\rho^2+\gamma_{ij}dx^idx^j\right)
\end{equation}
for $\ell$ AdS radius which we set to 1 for simplicity, $\rho$ holographic component using which we approach to the boundary $\partial \mathcal{M}$, and $\sigma=\pm1$ for (A)dS space. The metric on the boundary $\gamma_{ij}$ defines the first part of the boundary conditions in terms of the generalised Fefferman-Graham expansion
\begin{equation}
\gamma_{ij}=\gamma_{ij}^{(0)}+\rho\gamma_{ij}^{(1)}+\frac{1}{2}\rho^2\gamma_{ij}^{(2)}+.... \label{fg}
\end{equation} 
In addition, we take that variations of the first two terms in the expansion of the boundary metric (\ref{fg}) are 
\begin{align}
\delta \gamma_{ij}^{(0)}=\lambda\gamma_{ij}^{(0)}, & \delta \gamma_{ij}^{(1)}=2\lambda\gamma_{ij}^{(1)}.
\end{align}
%which define our boundary conditions. 
Since the metric $g_{\mu\nu}$ is invariant under small diffeomorphisms $x^{\mu}\rightarrow x^{\mu}+\xi^{\mu}$, its variation
\begin{equation}
\delta g_{\mu\nu}=\left(e^{2\omega}-1 \right)g_{\mu\nu}+\pounds_{\xi}g_{\mu\nu}\label{ke}
\end{equation} 
consists of the part that appears due to invariance under these diffeomorphisms and the part that appears due to invariance under Weyl rescalings. 
We expand the Weyl factor $\omega$ and the vector $\xi$  in the holographic coordinate $\rho$ analogously to the expansion of the boundary metric $\gamma_{ij}$  (\ref{fg}) and insert them in the Killing equation (\ref{ke}). This leads to equations that define the conditions on the components in the metric expansion (\ref{fg}). Taking $\delta g_{\rho\rho}=\delta g_{\rho i}=0$, we obtain that $\omega^{(0)}$ is zero and  $ij$ component of (\ref{ke}) defines the leading and the subleading order of the Killing equation. 
%In the equation (\ref{ke}) we insert the expansion of the Weyl factor $\omega$ in the 
The leading order Killing equation 
\begin{equation}
\mathcal{D}_i\xi^{(0)}_j+\mathcal{D}_j\xi^{(0)}_i=\frac{2}{3}\gamma_{ij}^{(0)}\mathcal{D}_k\xi^{(0)k}
\end{equation}
defines the leading term  in the expansion of the boundary metric, $\gamma_{ij}^{(0)}$, in the dependence on the leading order term in the expansion of Killing vectors (KV) $\xi$. The choice of the Minkowski metric for $\gamma_{ij}^{(0)}=\eta_{ij}=diag(-1,1,1)$ leads to vectors $\xi$ which define the conformal algebra $so(3,2)$ at the boundary. Conformal algebra is consisted of KVs of translations
%\begin{center}
\begin{align}
\xi^{(0)}&=\partial_t, & \xi^{(1)}&=\partial_x, & \xi^{(2)}&=\partial_y,
\end{align}
%\end{center}
Lorentz rotations \begin{align}
L_{ij}=x_i\partial_j-x_j\partial_i
\end{align}
%\end{center}
which define $\xi^{(3)}, \xi^{(4)}$ and $\xi^{(5)}$ for $i,j=t,x,y$,
the dilatation KV 
\begin{equation}
\xi^{(6)}=t\partial_t+x\partial_x+y\partial_y
\end{equation}
and special conformal transformations (SCTs) 
%\begin{center}
\begin{align}
\xi^{(7)}&=tx\partial_t+\frac{t^2+x^2-y^2}{2}\partial_x +xy\partial_y \\
\xi^{(8)}&=ty\partial_t+xy\partial_x+\frac{t^2+y^2-x^2}{2}\partial_y\\
\xi^{(9)}&=\frac{t^2+x^2+y^2}{2}\partial_t+tx\partial_x+ty\partial_y.
\end{align}
%\end{center}
We denote $\xi^t=\{\xi^{(0)},\xi^{(1)},\xi^{(2)}\}$ as translations, $\xi^{(6)}=\xi^d$ as dilatation, and $\xi^{sct}=\{\xi^{(7)},\xi^{(8)},\xi^{(9)}\}$ as SCT Killing vectors to define the $so(3,2)$
\begin{center}
\begin{align}
[\xi^d,\xi^t_i]&=-\xi^t_i && [\xi^d,\xi^{sct}_i]=\xi^{sct}_i\\
[\xi_k^t,L_{ij}]&=(\eta_{ki}\xi^t_j-\eta_{kj}\xi^t_i) && [\xi_{k}^{sct},L_{ij}]=-(\eta_{ki}\xi^{sct}_{j}-\eta_{kj}\xi^{sct}_i) \label{o321}
\end{align}
\begin{align}
[\xi_i^{sct},\xi_j^t]&=-(\eta_{ij}\xi^d-L_{ij})\\
[L_{ij},L_{kj}]&=-L_{ik}\label{o322}.
\end{align}
\end{center}
%The subleading order Killing equation further defines the
Due to the boundary conditions the subleading order Killing equation 
\begin{equation}
\pounds_{\xi^{(0)}}\gamma_{ij}^{(1)}=\frac{1}{3}\mathcal{D}_k\xi^{k}_{(0)}\gamma_{ij}^{(1)}\label{ske}
\end{equation}
consists of the Killing vectors $\xi_i^{(0)}$, leading term in the expansion of the boundary metric $\gamma_{ij}^{(0)}$, and the subleading term in the expansion of the boundary metric $\gamma_{ij}^{(1)}$.
From this equation we can proceed in the two possible directions. 
Since both $\xi_i^{(0)}$ and $\gamma_{ij}^{(1)}$ are undefined, we can choose the condition on the $\gamma_{ij}^{(1)}$, and inserting it in (\ref{ske}) determine the set of KVs $\xi_i^{(0)}$ which are conserved by that $\gamma_{ij}^{(1)}$. That set of KVs should also form a closed algebra. 
From the other side, we can choose the subalgebra of $so(3,2)$ defined by a set of $\xi^{(0)}_i$s %for which we want to determine the $\gamma_{ij}^{(1)}$, 
and find the corresponding $\gamma_{ij}^{(1)}$.

In this talk we focus on the approach in which, using (\ref{ske}) we determine $\gamma_{ij}^{(1)}$ for the chosen subalgebra of $so(3,2)$.
The $\gamma_{ij}^{(1)}$ can depend on all the coordinates on the boundary, while the simplest cases are of course those in which the components of the $\gamma_{ij}^{(1)}$ are constant. 

To be able to describe the required subalgebras, we will have to define the new KVs formed from the linear combinations of the existing ones. The most general linear combination of the KVs is
\begin{align}
\xi^{linear\text{ }combination}&=a_0\xi^{(0)}+a_1\xi^{(1)}+a_2\xi^{(2)}+a_3\xi^{(3)}+a_4\xi^{(4)}+a_5\xi^{(5)}+a_6\xi^{(6)}+a_7\xi^{(7)}\nonumber \\ &+a_8\xi^{(8)}+a_9\xi^{(9)} \label{lic}.
\end{align}
% When we insert the boundary conditions in the equation $\ref{ke}$ it is split in the components and provides a
In the following chapter we consider the largest subalgebras. % obtained in such a way.

\section{Asymptotic Symmetry Algebra of Conformal gravity}

The most interesting subalgebras of $so(3,2)$ for which we find $\gamma_{ij}^{(1)}$ are the highest dimensional subalgebras. They consist of five and four KVs.
%To classify them we have to examine the existence of $\gamma_{ij}^{(1)}$ for the highest subalgebras.
Here, we list first all the subalgebras of $so(3,2)$ \cite{Patera:1976my}
\begin{itemize}
\item similitude algebra, \textit{sim(2,1)}, 
\item optical algebra \textit{opt(2,1)},
\item maximal compact algebra $o(3)\otimes o(2)$
\item $o(2)\otimes o(2,1)$
\item \textit{o(2,2)}
\item Lorentz algebra $o(3,1)$
\item irreducible subalgbra $o(2,1)$,
\end{itemize}
while below we demonstrate how to define these subalgebras in terms of our KVs.

\subsubsection{Similitude algebra $sim(2,1)$}

The number of KVs in this algebra is 7, and we can  
%the boundary conditions allow as largest five dimensional ASA. 
 identify them with the KVs
\begin{align}
P_0&=\xi^{(0)}, && P_1=\xi^{(1)}, && P_{2}=\xi^{(2)} &&
F= \xi^{(6)} \\
 K_1&= \xi^{(3)} && K_2=\xi^{(4)} && L_3=\xi^{(5)}. && 
\end{align}
They close into \begin{align}
\left[\xi^d,\xi^t_j\right]&=-\xi^t_j \\
\left[\xi_l^t,L_{ij}\right]&=-\left(\eta_{li}\xi_j^t-\eta_{lj}\xi^t_i\right) \\
\left[L_{ij},L_{mj}\right]&=L_{im}.
\end{align}
If we insert the corresponding KVs in the subleading order Killing equation (\ref{ske}), we do not get a solution for $\gamma_{ij}^{(1)}$ unless it is trivial. The subalgebra of $sim(2,1)$ which contains five KVs is the highest dimensional one that leads to non-trivial $\gamma_{ij}^{(1)}$.

\subsubsection{Optical algebra $opt(2,1)$}

as well consists of 7 KVs which do not lead to non-trivial $\gamma_{ij}^{(1)}$. The realised subalgebra of $opt(2,1)$ is 5 dimensional.
When we write its KVs in a form
\begin{align}
W&=-\frac{\xi^{(6)}+\xi^{(4)}}{2} & K_1&=\frac{\xi^{(6)}-\xi^{(4)}}{2} & K_2&=\frac{1}{2}\left[\xi^{(0)}-\xi^{(2)}+\frac{(\xi^{(8)}-\xi^{(9)})}{2}\right] \\
 Q&=\frac{\xi^{(5)}-\xi^{(3)}}{2\sqrt2} & M&=-\sqrt{2}\xi^{(1)}&L_3&=\frac{1}{2}\left[\xi^{(0)}-\xi^{(2)}-\frac{(\xi^{(8)}-\xi^{(9)})}{2}\right]   \\
N&=-(\xi^{(0)}+\xi^{(2)}) & & &&
\end{align}
their algebra is defined with
\begin{align}
[K_1,K_2]&=-L_3,  & [L_3,K_1]&=K_2, & [L_3,K_2]&=-K_1, & [M,Q]&=-N, & [K_1,M]&=-\frac{1}{2} M, \\ [K_1,Q]&=\frac{1}{2}Q, & [K_1,N]&=0, &[K_2,M]&=\frac{1}{2}Q, &
 [K_2,Q]&=\frac{1}{2}M, & [K_2,N]&=0   \\
[L_3,M]&=-\frac{1}{2}Q, & [L_3,Q]&=\frac{1}{2}M, & [L_3,N]&=0 &
[W,M]&=\frac{1}{2}M,  & [W,Q]&=\frac{1}{2}Q, \\ [W,N]&=\frac{1}{2}N.
  \end{align}

\subsubsection{$o(2,2)$}

The algebra $o(2,2)$ is 6 dimensional while its highest realised subalgebra consists of 4 KVs.
Rewriting the KVs in the form,
 \begin{align}
A_1&=-\frac{1}{2}\left[ \frac{\xi^{(9)} + \xi^{(8)} }{ 2}  -\left( \xi^{(0)} +\right) \right],  & A_2 &=\frac{1}{2} \left(\xi^{(6)} +\xi^{(4)} \right), \\
A_3 & =\frac{1}{2}\left[ -\frac{\xi^{(9)}+\xi^{(8)}}{2} -\left( \xi^{(0)} +\xi^{(2)}\right)\right]  & B_1&=-\frac{1}{2}\left[\frac{-\xi^{(9)}-\xi^{(8)}}{2} +\left( \xi^{(0)}-\xi^{(2)}\right) \right] \\ 
B_2&=\frac{1}{2} \left( \xi^{(6)}-\xi^{(4)}\right) &B_3&=\frac{1}{2}\left[  \frac{\xi^{(9)}-\xi^{(8)}}{2}+\left(\xi^{(0)} -\xi^{(2)}\right)\right] 
\end{align}
we can identify the algebra
\begin{align}
[A_1,A_2]&=-A_3 & [A_3,A_1]&=A_2 & [A_2,A_3]&=A_1 \\
[B_1,B_2]&=-B_3 & [B_3,B_1]&=B_2 & [B_2,B_3]&=B_1 
\end{align}
which can be summarised into $ [A_i,B_k]=0 $ for $(i,k)=1,2,3$.

%we can define its algebra as

\subsubsection{$o(3,1)$} 

This algebra consists of 6 KVs, while the highest dimensional realised subalgebra is 4 dimensional.
%The $o(3,1)$
If we define the KVs as
\begin{align}
L_1&=\xi^{(7)}+\frac{\xi^{(2)}}{2}  & L_2& =\xi^{(5)}  & L_3&=\xi^{(8)}+\frac{1}{2}\xi^{(1)} \\
K_1 &=\xi^{(8)}-\frac{1}{2}\xi^{(1)} & K_2&=\xi^{(6)} & K_3&=-\xi^{(7)}+\frac{1}{2} \xi{(2)}
\end{align}
 the algebra is
\begin{align}
\left[L_i,L_j\right]&=\epsilon_{ijk}L_k, \\
\left[L_i,K_j  \right]&=\epsilon_{ijk}K_k, \\
\left[K_i,K_j\right]&=-\epsilon_{ijk}L_k.
\end{align}
The subalgebras which have for the highest  dimensional  realised  subalgebra, algebra with 4 KVs are $o(3)\otimes o(2)$, $o(2)\otimes o(2,1)$ and $o(2,1)$ and we will not consider their algebras and KVs explicitly here, for more information see \cite{Irakleidou:2016xot}.
 The explicit solutions defined by some of the algebras we mentioned above are written in the table in the Appendix: "Highest realised subalgebras of $sim(2,1)$", while now we focus on the  realised subalgbras that can be extended to global solutions.

\section{Global Solution}

To obtain the  five dimensional subalgebra we consider the KVs of the similitude algebra which form the subalgebra $\alpha_{5,4}$ (adopting the notation from \cite{Patera:1976my}) and insert them in the equation (\ref{ske}). The equation (\ref{ske}) is solvable when $\gamma_{ij}^{(1)}$
 is 
 \begin{align}
 \gamma_{ij}^{(1)}&=\left( \begin{array}{ccc}  c & c& 0  \\ c& c& 0 \\ 0& 0& 0\end{array} \right),\label{five}
 \end{align}
 and the KVs that define $\alpha_{5,4}$ are KVs of translation and two new linearly combined KVs
 \begin{align}
\chi_{new}^{(1)}&=\xi^{(6)}-\frac{1}{2}\xi^{(3)} & \chi_{new}^{(2)}&=\xi^{(5)}-\xi^{(4)}.
\end{align}
 The solution (\ref{five}) helps us to find the global soultion, because we can use it to write the ansatz metric
 \begin{align}
 ds^2=dr^2+(-1+cf(r))dx_i^2+2cf(r)dx_idx_j+(1+cf(r))dx_j^2+dx_x^2\label{ansatz}
 \end{align}
 for $x_i=\{t,x,y\}$.
 The line element (\ref{ansatz}) gives global solution and solves Bach equation (\ref{bach}) for $f(r)=c_1+c_2+c_3 r^2+c_4 r^3$.
 
Interestingly, we can notice form the solution for $f(r)$ that choosing the  coefficients $c_i$ ($i=1,2,3,4$) we can decide whether we will have the corresponding charges. The AdS/CFT correspondence tells us how to define the stress energy tensors at the boundary, and for conformal gravity and this particular metric, stress energy tensors are partially massless response $P_{ij}$ in the sense of Deser, Nepomechie, and Waldron \cite{Deser:2012qg, Deser:2013uy}, and standard Brown-York stress energy tensor $\tau_{ij}$.
We  obtain
\begin{align}
\tau_{ij}=\left(  \begin{array}{ccc}-c c_4 & -c c_4 &0 \\ -cc_4 &-cc_4 &0 \\ 0& 0&0 \end{array} \right)
\end{align}
and \begin{align}
P_{ij}=\left(  \begin{array}{ccc} c c_3 & c c_3 &0 \\ cc_3 &cc_3 &0 \\ 0& 0&0 \end{array} \right).
\end{align}
Using the analogous ansatz metric and $\gamma_{ij}^{(1)}$ of the form
\begin{align}
\gamma_{ij}=\left(  \begin{array}{ccc} -c b(t-y) & 0 & c b(t-y) \\ 0&0  &0 \\ c b(t-y)& 0& -cb(t-y) \end{array} \right).
\end{align}
which conserve the KVs 
\begin{align}
\xi^{(n1)}&=-P_0+P_2 & \xi_2^{(n2)}&=P_1 & \xi^{(n3)}&=P_1 &\xi^{(n3)}&=L_3-K_1
\end{align}
we as well obtain the global solution of the Bach equation (\ref{bach}).

The particularity of this  metric is that $b(t-y)$  is a function, which allows us to solve (\ref{ske}) for the further conserved KV.  We can obtain the forms of $\gamma_{ij}^{(1)}$ for the corresponding KV written in the table
\begin{center}
\begin{tabular}{|c|c|c|c|}
%\caption{Table 1}
%\begin{tabular}{ |  p{4 cm} | p{5.7 cm} || p{4cm}|} | b(t-y) |
\hline
 4th KV & b(t-y)  & 4th KV & $b(t-y)$  \\ \hline
$F$ & $\frac{b}{t-y}$ & $F-K_2$ &  $\frac{b}{(t-y)^{3/2}}$ \\ \hline
$F+K_2+\epsilon(-P_0-P_2)$ & $b\cdot e^{\frac{t-1}{2\epsilon}}$ & $K_2$ & $\frac{b}{(t-y)^2}$ \\ \hline
$P_0-P_2$ & b(t-1) & $F+c K_2$ & $b\cdot (t-y)^{\frac{1-2c}{-1+c}}$.\\
\hline
\end{tabular}
\end{center}
%\begin{table}

%\end{table}

\section{Conclusion and Outlook}

Depending on the linear term in the FG expansion, we can impose a number of boundary conditions in conformal gravity.
These boundary conditions are classified according to subalgebras of $so(3,2)$, and with the clever choice of an ansatz metric, using the realised $\gamma_{ij}^{(1)}$ matrices, we can obtain a global solution. 
Global solutions can therefore be classified according to the realised subalgebras.

Largest asymptotic symmetry algebra we find is 5 dimensional, belongs to subalgbra of $sim(2,1)$ and $opt(2,1)$ and defines pp wave or geon solution.
$o(2,2)$ and $o(3,1)$ algebras define ASAs with maximally 4 KVs. 
Each of these can also give pp wave global solution.%, while further global solutions we comment in \cite{Irakleidou}.

However, there are more global solutions. To find them using this approach, one has to carefully choose the ansatz metric and depending on the requirement of the global solutions, impose additional conditions which may  lead to easy or demanding task.
Further research in this direction include studying the black hole solutions, black branes and black strings.

\section{Acknowledgements} 
 I would like to thank the organisers of the "School and Workshops on Elementary Particle Physics and Gravity" at Corfu Summer Institute 2016 for the hospitality. The work was funded by the {\it Forschungsstipendien 2015}  of {\it Technische Universit\"at Wien}.

\section{Appendix: Highest realised subalgebras of $sim(2,1)$}

Here we write an example of the asymptotic solutions for $\gamma_{ij}^{(1)}$ 
and their corresponding algebra. The subalgebras of $sim(2,1)$ we denote with "$\alpha$" and adopt the notation of \cite{Patera:1976my}, while the generators are identified with with $
P_0=-\xi^{(0)},P_1=\xi^{(1)},P_2=\xi^{(2)},F=\xi^{(6)},K_1=\xi^{(3)}, K_2=\xi^{(4)},L_3=\xi^{(5)}.%\label{simid}
$

%\begin{center}
%\begin{table}[ht!]
%\caption{ Realized subalgebras, sim(2,1)}
\hspace{-0.5cm}\begin{tabular}{ | l  | p{2.5 cm} | p{4,0 cm} | p{6.0cm} |}
 \hline
  \multicolumn{4}{|c|}{Realized subalgebras} \\
  \hline
 Name &Patera name&generators  & $\gamma_{ij}^{(1)}$  \\ \hline\hline
&$\alpha_{5,4}^a$ & $\begin{array}{l}F+\frac{1}{2}K_2,-K_1+L_3,\\P_0,P_1,P_2\end{array}$  & $\gamma_{ij}^{(1)}=\left(
\begin{array}{ccc}
 -b & 0 & b \\
 0 & 0 & 0 \\
 b & 0 & -b \\ 
\end{array}
\right)$
\\
&$\alpha\neq0,\pm1$&$\alpha=\frac{1}{2}$&\\
$\mathcal{R}\oplus o(3)$ &$\alpha_{4,1}$ & $P_1\oplus\left\{K_2,P_0,P_2\right\}$  & $ \left(\begin{array}{ccc} \frac{b}{2} & 0 & 0 \\ 0 & b &0 \\ 0 & 0 & -\frac{b}{2}  \end{array}\right) $ \\
 &$\alpha_{4,2}$ & $\begin{array}{l}P_0-P_2\oplus\\ \left\{F-K_2;P_0+P_2,P_1\right\}\end{array}$  &  $  \left(\begin{array}{ccc} -b & 0 & 0 \\ 0 & b &0 \\ 0 & 0 & -2b   \end{array}\right)$\\ $\begin{array}{c}
MKR \\  \mathcal{R}\oplus o(3)\end{array}$ &$\alpha_{4,3}$ & $P_0\oplus\left\{L_3,P_1,P_2\right\}$  & $  \left(\begin{array}{ccc} 2b & 0 & 0 \\ 0 & b &0 \\ 0 & 0 & b   \end{array}\right) $ \\ &$\alpha_{4,4}$ & $F\oplus\left\{K_1,K_2,L_3\right\}$ & $
\gamma_{ij}^{(1)}=\left(\begin{array} {ccc} 2f(t) & 0 & 0 \\ 0 & f(t) &0 \\ 0 & 0 & f(t)  \end{array}\right)$
\\ &$\alpha_{4,5}$ & $\begin{array}{l}F\{K_2;P_0-P_2\}\oplus \\ \left\{F-K_2,P_1\right\}\end{array}$ &$ \left(
\begin{array}{ccc}
 0 & \frac{b}{t-y} & 0 \\
 \frac{b}{t-y} & 0 & \frac{b}{y-t} \\
 0 & \frac{b}{y-t} & 0 \\
\end{array}
\right)$\\
%%%
&$\alpha_{4,6}$ & 
$ \begin{array}{c} \left\{ F+K_2,P_0-P_2 \right\}\oplus  \\ \left\{F-K_2,P_0+P_2\right\} \end{array} $
 &
$\left(
\begin{array}{ccc}
 \frac{b}{x} & 0 & 0 \\
 0 & \frac{2 b}{x} & 0 \\
 0 & 0 & -\frac{b}{x} \\
\end{array}
\right)$%,and (\ref{poind}) 
\\
&$\alpha_{4,7}$&$ \begin{array}{c}  L_3-K_1,P_0+P_2 ;  \\ P_0-P_2,P_1 \end{array} $& gives 5 KV subalgebra for constant components of $\gamma_{ij}^{(1)}$\\
\hline 
 \end{tabular}
 %\label{tablesim}
 %\end{table}
%\end{center}

%\bibliography{bibliothek}

%\bibliography{bibliothek1}

\end{document}